\documentstyle[aps]{revtex}
\begin{document}
\twocolumn[
{\bf Comment on ``Dimensional and dynamical aspects of the Casimir
effect: understanding the reality and significance of vacuum
energy''}\\
\smallskip ]

In \cite{KM00} Milton has presented a brief review of some interesting
aspects of the theory of the Casimir effect.  This comment is
aimed at correcting some imprecise statements in that work with
respect to the relevance of the Casimir effect for explaining the
phenomenon of sonoluminescence. In particular, \cite{KM00} makes three
assertions with which we strongly disagree
\begin{enumerate}
\item 
That all of the Casimir-based explanations of sonoluminescence rely on
static Casimir effect calculations.
\item That the proposal by Liberati {\emph{et
      al.}}~\cite{SnPr,PRL,Qed1,Qed2,2Gamma} is based on the same
  mechanism as used by Schwinger~\cite{Sch} (photon production from
  the quantum vacuum generated by the rapid movement of the bubble
  boundary).
\item 
That the ``temperature'' of the photons emitted in
sonoluminescence can be identified with the Unruh temperature
estimated from the acceleration of the dielectric boundary.
\end{enumerate}
Let us explain the motivation behind our comments:\\

{\bf Relevance of the static Casimir effect}: There are at least three
models which apply the {\emph{dynamical}} Casimir effect (particle
production from the quantum vacuum by a time-varying external field)
to the problem of sonoluminescence. In particular Eberlein~\cite{Ebe},
Sch\"utzhold {\emph{et al.}}~\cite{SPS}, and the main part of the present
authors' work~\cite{PRL,Qed1} make explicit use of such a framework
(although the ``external field'' driving the particle creation is not
the same in all of these models). It is then hard to agree with the claims
made in \cite{KM00} that ``the dynamical Casimir effect remains
largely unknown'' and that ``Schwinger and his followers had to rely
on the known results for the Casimir effect with static boundary
conditions''.

{\bf Schwinger vs.\ Liberati et al. proposal}: The proposal by
Liberati et al.~\cite{SnPr,PRL,Qed1,Qed2,2Gamma} cited in section 4
of~\cite{KM00} is not at all coincident with the Schwinger
proposal~\cite{Sch}.  Neither do the objections raised by Milton to
the latter proposal apply straightforwardly to the former. In
particular, it is not true that we proposed a model based on the
instantaneous collapse of a bubble: Such a calculation is performed
in~\cite{Qed0} as a check of Schwinger's proposal. It is there clearly
shown that, although correct in principle, Schwinger's mechanism
cannot be applied to sonoluminescence due to the extremely short
timescales (femtoseconds) required for the collapse.  It is not the
femtosecond scale in itself that is unphysical ({\emph{e.g.}}
femtosecond lasers are now well established technology), simply that
the known timescale for collapse of the sonoluminescent bubble is not
so short.

In view of this, we proposed a modified mechanism where particle
production from the QED vacuum is due to a sudden change in the
refractive index of the gas. In our model the bubble radius is taken
to be at its minimum (its changes are presumably non-relativistic even
near to the end of the collapse), whereas the refractive index of the gas
inside the bubble is assumed to be changing rapidly.  In~\cite{Qed1}
we calculated the number and the energy of the photons that can be
produced by a {\em time-dependent refractive index} of the gas at
fixed bubble radius. We also demonstrated that a timescale {\em up to
  the order of picoseconds} could be sufficient to explain the
observed emission (depending on the initial and final values of the
refractive index), thus we are allowed to avoid any superluminal
propagation of signal inside the bubble.  The semi-static
approximation (instantaneous change in the bubble refractive index),
which Milton seems to view as pre-eminent, instead appears only
in~\cite{Qed2} and then only as a consistency check used to extend and
confirm the results of the companion paper~\cite{Qed1} to the case in
which finite volume effects are taken into account.

{\bf Unruh temperature}: A fundamental point is central to all of the
vacuum-based explanations of sonoluminescence: In these models nothing
inside the bubble reaches the physical temperature of {\em thousands
  of Kelvin}.  A thermal spectrum (even one at very large temperature)
can be {\emph{simulated}} by particle states generated by the dynamical
Casimir effect, as is well-known and explained in an extensive literature
({\emph{e.g.}} see~\cite{2Gamma} and references therein).  Such a
simulation of thermality requires a particular regime~\cite{2Gamma} and
is thus {\emph not} generic.  [Notice that it is controversial
whether a truly thermal spectrum is actually detected at all in
sonoluminescence.]  Moreover the Unruh temperature $T=a/2\pi$ is pertinent
to the case of the response in a Minkowski vacuum of a
detector endowed with uniform acceleration $a$~\cite{Unruh}. It appears
very hard to apply this situation to a collapsing bubble (even in the
case where one uses the moving bubble boundary as the external field
driving quantum particle creation~\cite{Ebe,SPS}).

In view of the above facts, to require the Unruh temperature to be
several thousand Kelvin, and from this to deduce the necessary
timescale for the bubble collapse, is simply an incorrect
procedure. The fact that following such a procedure Milton arrives at
a result in agreement with our conclusions presented in~\cite{Qed0} is
mainly due to a common feature of all of the dynamical Casimir effects:
the typical timescale required for the change of the external field is
roughly inversely proportional to the typical frequency of the
particles which one wants to produce. In order to justify a spectrum peaked
at frequencies of the order of the Petahertz (which in
sonoluminescence is often interpreted as a thermal spectrum at several
thousand Kelvin) one apparently needs timescales of about one
femtosecond. The reason why this does not apply to our model is that
the typical frequencies of the particles produced by a time-varying
refractive index are {\em not} just determined by the inverse of the
typical timescale of this variation but also by the initial and final
values of the refractive index.

In conclusion we hope that this ``comment'' will serve to remind
people that the Casimir-based explanation of sonoluminescence cannot
be dismissed out of hand.  It is possible that experiments will
ultimately disprove the dynamical Casimir effect based explanation for
sonoluminescence, but this will not happen on the basis of the
arguments proposed in~\cite{KM00}.

\bigskip
\noindent
Stefano Liberati\\
SISSA/ISAS\\
Via Beirut 2-4, 34014 Trieste.\\
e-mail: liberati@sissa.it\\

\noindent
Francesco Belgiorno\\
Universit\`a di Milano,\\
Via Celoria 16, 20133 Milano.\\
e-mail: belgiorno@mi.infn.it\\

\noindent
Matt Visser\\
Washington University in Saint Louis\\
Saint Louis, Missouri 63130-4899\\
e-mail: visser@kiwi.wustl.edu\\
%
%

\end{document}